\documentclass[conference]{IEEEtran}
\IEEEoverridecommandlockouts
\usepackage{cite}
\usepackage{amsmath,amssymb,amsfonts}
\usepackage{algorithmic}
\usepackage{graphicx}
\usepackage{textcomp}
\usepackage{xcolor}

\def\BibTeX{{\rm B\kern-.05em{\sc i\kern-.025em b}\kern-.08em
    T\kern-.1667em\lower.7ex\hbox{E}\kern-.125emX}}

\usepackage[hyphens]{url}
\usepackage{balance}
\usepackage[linesnumbered,ruled,vlined]{algorithm2e}
\usepackage{multirow}
\usepackage{array}
\usepackage{varioref}
\usepackage{amssymb}
\usepackage{pifont}
\usepackage{tikz}
\usepackage{booktabs}
\usepackage{balance}
\usepackage{hyperref}

\usetikzlibrary{matrix,chains,positioning,decorations.pathreplacing,arrows,backgrounds}
\usepackage{float}
\usepgflibrary{shapes.multipart}
\usepackage{tikz}
\usepackage{pgfplots}
\usepackage{subfigure}
\usepackage{pgf-umlsd}
\usepackage{footnote}
\usepackage[most]{tcolorbox}
\usepackage[export]{adjustbox}

\usepackage{color, colortbl}
\definecolor{Gray}{gray}{0.9}
\definecolor{airforceblue}{rgb}{0.36, 0.54, 0.66}
\definecolor{aliceblue}{rgb}{0.94, 0.97, 1.0}
\definecolor{alizarin}{rgb}{0.82, 0.1, 0.26}
\definecolor{amber}{rgb}{1.0, 0.75, 0.0}
\definecolor{amber(sae/ece)}{rgb}{1.0, 0.49, 0.0}
\definecolor{awesome}{rgb}{1.0, 0.13, 0.32}
\definecolor{babypink}{rgb}{0.96, 0.76, 0.76}
\definecolor{bronze}{rgb}{0.8, 0.5, 0.2}
\definecolor{battleshipgrey}{rgb}{0.52, 0.52, 0.51}
\definecolor{bole}{rgb}{0.47, 0.27, 0.23}
\definecolor{bulgarianrose}{rgb}{0.28, 0.02, 0.03}
\definecolor{brinkpink}{rgb}{0.98, 0.38, 0.5}
\definecolor{cadet}{rgb}{0.33, 0.41, 0.47}
\definecolor{ceil}{rgb}{0.57, 0.63, 0.81}
\definecolor{cerulean}{rgb}{0.0, 0.48, 0.65}
\definecolor{charcoal}{rgb}{0.21, 0.27, 0.31}
\definecolor{coolblack}{rgb}{0.0, 0.18, 0.39}
\definecolor{coolgrey}{rgb}{0.55, 0.57, 0.67}
\definecolor{darkcandyapplered}{rgb}{0.64, 0.0, 0.0}
\definecolor{darkbrown}{rgb}{0.4, 0.26, 0.13}
\definecolor{darkcerulean}{rgb}{0.03, 0.27, 0.49}
\definecolor{darkgray}{rgb}{0.66, 0.66, 0.66}
\definecolor{darkgoldenrod}{rgb}{0.72, 0.53, 0.04}
\definecolor{darkjunglegreen}{rgb}{0.1, 0.14, 0.13}
\definecolor{darktaupe}{rgb}{0.28, 0.24, 0.2}
\definecolor{davy\'sgrey}{rgb}{0.33, 0.33, 0.33}
\definecolor{frenchblue}{rgb}{0.0, 0.45, 0.73}
\definecolor{almond}{rgb}{0.94, 0.87, 0.8}
\definecolor{beaublue}{rgb}{0.74, 0.83, 0.9}
\definecolor{beige}{rgb}{0.96, 0.96, 0.86}
\definecolor{bisque}{rgb}{1.0, 0.89, 0.77}
\definecolor{black}{rgb}{0.0, 0.0, 0.0}
\definecolor{fluorescentorange}{rgb}{1.0, 0.75, 0.0}
\definecolor{ghostwhite}{rgb}{0.97, 0.97, 1.0}
\definecolor{antiquewhite}{rgb}{0.98, 0.92, 0.84}
\definecolor{ao(english)}{rgb}{0.0, 0.5, 0.0}

\newcommand{\TitleFull}{Blind Faith}

\begin{document}
	

\title{\TitleFull: Privacy-Preserving Machine Learning using Function Approximation 
{\footnotesize \textsuperscript	\hfill\href{https://zenodo.org/deposit/5143787}{\includegraphics[width=2.5\baselineskip,height=3\baselineskip]
		{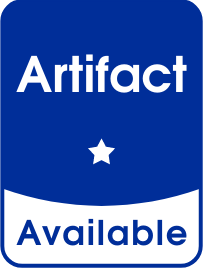}}}
\thanks{Funded by the ASCLEPIOS: Advanced Secure Cloud Encrypted Platform for Internationally
Orchestrated Solutions in Healthcare Project No. 826093 EU research project.}
}

\author{\IEEEauthorblockN{1\textsuperscript{st} Tanveer Khan }
\IEEEauthorblockA{\textit{Department of Computing Sciences,} \\
\textit{Tampere University,}\\
Tampere, Finland \\
tanveer.khan@tuni.fi}
\and
\IEEEauthorblockN{2\textsuperscript{nd} Alexandros Bakas}
\IEEEauthorblockA{\textit{Department of Computing Sciences,} \\
\textit{Tampere University,}\\
Tampere, Finland \\
alexandros.bakas@tuni.fi}
\and
\IEEEauthorblockN{3\textsuperscript{rd} Antonis Michalas }
\IEEEauthorblockA{\textit{Department of Computing Sciences,} \\
\textit{Tampere University,}\\
Tampere, Finland \\
antonios.michalas@tuni.fi}
}

\maketitle

\begin{abstract}
Over the past few years, a tremendous growth of machine learning was brought about by a significant increase in adoption of cloud-based services. As a result, various solutions have been proposed in which the machine learning models run on a remote cloud provider. However, when such a model is deployed on an untrusted cloud, it is of vital importance that the users' privacy is preserved. To this end, we propose Blind Faith -- a machine learning model in which the training phase occurs in plaintext data, but the classification of the users' inputs is performed on homomorphically encrypted ciphertexts. To make our construction compatible with homomorphic encryption, we approximate the activation functions using Chebyshev polynomials. This allowed us to build a privacy-preserving machine learning model that can classify encrypted images. \TitleFull{} preserves users' privacy since it can perform high accuracy predictions by performing computations directly on encrypted data.
\end{abstract}


\section{Introduction}
\label{sec:intro}
Machine Learning (ML) models have attracted global adulation and are used in a plethora of applications such as medical diagnosis, pattern recognition, and credit risk assessment. Recently, deep learning -- a sub-field of ML has gained extra attention from researchers due to its solid performance in many tasks such as speech recognition, spam detection, image classification, traffic analysis, face recognition, financial detection and genomics prediction~\cite{islam2011application, kim2015private, gilad2016cryptonets, dowlin2017manual, shortell2019secure}. ML models normally consist of a training and a testing phase. In the training phase, the model is trained on a dataset while in the testing phase, the model predicts the output using a previously unseen input. Due to the increased demand for these services, Cloud Service Providers (CSP) also offer Machine Learning as a Service (MLaaS). In MLaaS, the training and deploying of ML models are performed using the CSP infrastructure. Once the models are trained and deployed on the CSP, the users can use these models for online prediction services. MLaaS have clear benefits and such services are currently offered  by providers including Google Prediction API~\cite{Google}, Microsoft Azure ML~\cite{Microsoft}, and Ersatz Lab~\cite{Ersatz}. The problem with MLaaS is that either the training or testing, or both phases, need to be outsourced to the CSP. The outsourced data can be of a sensitive nature, as for example in the areas of healthcare and finance~\cite{Michalas:14:Healthcom}, which, in its turn, raises privacy concerns. As an example, the data sent to the prediction models may be misused or stolen. To preserve the privacy of user data, different methods have been proposed~\cite{Patricia, garge2018neural, gilad2016cryptonets}.

The purpose of this work is to illustrate the use of Neural Networks (NN) over encrypted data. We use Homomorphic Encryption (HE) to perform computations on encrypted data. HE can perform arithmetic operations (addition and multiplication) over encrypted data without decrypting it. Therefore, any function that uses these arithmetic operations can be homomorphically evaluated. Furthermore, we consider the Convolutional Neural Network (CNN) to be homomorphically evaluated on the encrypted data. All the operations except activation functions in CNN are simple addition and multiplication and can be homomorphically evaluated. The activation functions such as rectified-linear unit (ReLU), Sigmoid, etc. are non-linear and cannot be evaluated homomorphically.  

For CNN to be homomorphically evaluated on encrypted data, designing efficient HE-friendly support for activation function has been an active topic of research. Different methods are proposed to support the non-linear activation functions such as power functions~\cite{gilad2016cryptonets}, look-up table~\cite{crawford2018doing} and polynomial approximations~\cite{hesamifard2016cryptodl, chabanne2017privacy, chou2018faster}. In this work, we use low degree Chebyshev polynomials to approximate a non-linear activation function.

Chebyshev polynomials allow us to efficiently compute any continuous function in a given interval, using only low degree polynomials. This is an important feature as our computations become significantly more efficient and lower the overall computational complexity.

\subsection{Our Contribution}
\label{subsec:our contribution}
The contributions of this paper are manifold. First, we show how to approximate activation functions using Chebyshev polynomials. As a next step, we used the approximated functions to design an NN model. Moreover, we combined our design with HE in order to deploy \TitleFull{} -- a privacy-preserving ML (PPML) model that can classify encrypted images with high accuracy. Our model has a wide application domain since it can be applied to fields where users' data is considered sacrosanct. For example, consider a cloud-based eHealth service where patients can upload an encrypted medical image and the cloud can approximately tell if the user suffers from a certain disease or not without getting any information about the actual user, the data sent to the cloud or any personal medical records. 
To illustrate the effectiveness of our model, we conducted extensive experiments. 
Our contributions can be summarized as follows:

\begin{itemize}
	\item We approximate the ReLU and Sigmoid activation functions using Chebyshev polynomials and compare the differences in terms of efficiency and accuracy.
	\item We design a PPML model in which the CNN is trained on plaintext data while the classification process is performed on homomorphically encrypted data.
	\item We compare our design with other state-of-the-art works in the field of PPML.
	\item Our construction obviates the need of a powerful machine on the client's side as the classification takes place on the CSP. Hence, the role of client can be played by a constrained device whose sole responsibility will be to encrypt an outsource a file to the CSP.
\end{itemize}

\subsection{Organization}
\label{subsec:organization}
The rest of the paper is organized as follows: In Section~\ref{sec:relatedwork}, we present important published works in the area of PPML. 
In Section~\ref{sec:preliminaries}, we show how to approximate the ReLU and Sigmoid activation functions using low degree Chebysev polynomials. The methodology of our work is illustrated in Section~\ref{sec:methodology}, which is followed by extensive experimental results in Section~\ref{sec:performance analysis} and finally, in Section~\ref{sec:Impact} we conclude the paper.

\section{Related Work}
\label{sec:relatedwork}
The first steps towards the development of a PPML model were implemented using Multiparty Computation (MPC). In an MPC scheme, the parties jointly compute a function while keeping the original inputs private. A plethora of different methods based on MPC has been proposed as a way to preserve the privacy of the underlying ML models~\cite{bunn2007secure, lindell2000privacy, vaidya2008privacy}. 

Mohassel \textit{et al.,}~\cite{mohassel2017secureml} designed SecureML -- an efficient protocol for preserving the privacy of various ML models using MPC. This protocol is based on a two-server model, where the data owner first distributes the data among two \textit{non}-colluding servers. These servers then train various models on the joint data using secure MPC. In addition, the protocol provides support to approximate the activation function during the training phase. Since SecureML requires changes in the training phase, the model does not apply to the problem of making the existing NN model oblivious. MiniONN~\cite{liu2017oblivious} is another MPC-based approach used to convert any NN to an oblivious NN supporting privacy-preserving predictions with reasonable efficiency. Oblivious protocols are designed to minimize the accuracy loss in linear transformation, pooling operations and activation functions. While MiniONN uses lightweight cryptographic primitives, such as garbled circuits and secret sharing, it still reveals information about the network (e.g. size of the filter)~\cite{juvekar2018gazelle}. 
%
%
Wu \textit{et al.}~\cite{wu2013privacy} proposed a privacy-preserving logistic regression model. As the logistic function is \textit{not} linear, the authors use polynomial fitting to achieve a good approximation. However, by doing so the accuracy of the model is degraded. Graepel \textit{et al.}~\cite{graepel2012ml} used Somewhat Homomorphic Encryption (SHE)~\cite{fan2012somewhat} to train two simple classifiers: Linear Mean classifier~\cite{duda2006pattern} and Fisher's Linear Discriminate classifiers~\cite{fisher1936use} on encrypted data. To perform efficient computations on encrypted data, low degree polynomials were deployed. The proposed approach does not offer strong security guarantees as the model is leaked to the users. 

Privacy-preserving classification for Naive Bayes, Hyperplane Decision, and Decision tree was provided by Bost \textit{et al.}~\cite{bost2015machine} using a combination of HE and garbled circuits. The basic idea was to provide users with the encrypted classification parameters and to compute the prediction values using a partially HE scheme~\cite{paillier1999public}. However, this approach fails to preserve users' privacy. This has been demonstrated in~\cite{gao2018privacy}, where a \textit{substitution-then-comparison} attack, is enough to reveal \textit{all} information of one party to another.

A design based on Leveled HE (LHE)~\cite{brakerski2014leveled} is presented by Ehsan \textit{et al.}~\cite{hesamifard2017cryptodl}. Their goal was to utilize the properties of the LHE scheme to preserve the privacy of CNN while at the same time keep the accuracy as close as possible to the original model. In their work, the authors approximated the Sigmoid, ReLU and Tanh activation functions and achieved an accuracy of~99.52\% on the MINST dataset~\cite{lecun-mnisthandwrittendigit-2010}. This is a remarkable result, as the accuracy of the original model was measured at~99.56\%. Unfortunately, their approach is computationally expensive, as both the training and testing phases are performed on encrypted data.


\paragraph{\textbf{Most Relevant Related Work:}} In~\cite{gilad2016cryptonets}, authors proposed CryptoNets -- an NN model applied to encrypted data. While CryptoNets achieves remarkable accuracy, the construction is based on the use of square activation function. Hence, approximating a non-linear function causes instability during the training phase when the interval\footnote{By interval we mean the domain of definition of the activation function.} is large. In our work, we overcome this problem by using Chebyshev approximation. More specifically, Chebyshev polynomials allows to approximate the activation function accurately even in larger intervals. Using Chebyshev polynomials, is \textit{not} a novel idea. In~\cite{hesamifard2018privacy}, authors proposed a framework in which both the training and the classification phases are performed over encrypted data. In this work, authors used the Chebyshev polynomials to approximate the derivative of the activation functions. While their results are quite impressive, the classification process is computationally expensive. This is due to the fact that the ML model is encrypted and hence, classification is performed on encrypted data. To achieve better efficiency, we adapted a hybrid approach in which the client's input is encrypted but the model is in plaintext.

\section{Preliminaries}
\label{sec:preliminaries}
\subsection{Homomorphic Encryption}
\label{subsec:HE}

An HE scheme consists of the four following algorithms:

\begin{itemize}
	\item $\mathsf{KeyGen}$: A randomized algorithm that on input a security parameter $\lambda$, outputs a public/private key pair $\mathsf{(pk, sk)}$.
	
	\item $\mathsf{Enc}$: A randomized algorithm that on input a message $m$ and a public key $\mathsf{pk}$, outputs a ciphertext $c$.
	
	\item $\mathsf{Dec}$: A deterministic algorithm that on input a ciphertext $c$ and a secret key $\mathsf{sk}$, outputs a plaintext $m$.
	
	\item $\mathsf{Eval}$: A randomized algorithm that on input a public key $\mathsf{pk}$, a function $f$, and a set of ciphertexts $c_1, \dots, c_n$, outputs $f(c_1, \dots, c_n)$.
	
\end{itemize}

\subsection{Background on Polynomial Approximations}
\label{subsec:polynomial approximation}
Approximating continuous functions is a problem that has drawn mathematicians' attention for a very long time. While there are several ways to approximate a continuous function, in this work we are only interested in polynomial approximations. More specifically, we are using Chebysev polynomials to approximate the Sigmoid and the ReLU functions. However, there are various works that use different approaches such as the $x^2$ function~\cite{gilad2016cryptonets}, the \textit{Piecewise} approximation~\cite{chabanne2017privacy}, lookup tables~\cite{crawford2018doing} etc. Unfortunately, all these methods face certain limitations. For example, the $x^2$ method can cause instability during the training phase and the creation of a piecewise linear approximation can sometimes be a complex optimization problem. With this in mind, we chose to work with Chebyshev polynomials. 

\begin{equation}
	\label{equ:chebyshev}
	T_{n+1}(x)=2xT_{n}(x)-T_{n-1}(x)
\end{equation}

\subsection{Chebyshev Polynomials}
\label{subsec:chebyshev approximation}

In this Section, we show how low degree Chebyshev polynomials can be utilized to approximate the activation functions. 
%
Chebyshev approximation is also known as the minimax approximation.  The minimax polynomial approach is used for function approximation by improving the accuracy and lowering the overall computational complexity~\cite{schlessman2002approximation}. Instead of minimizing the error at the point of expansion like Taylor's polynomial approximation, the minimax approach minimizes the error across a given input segment. The minimax approximation is used to find a mathematical function that minimizes the maximum error. As an example, for a function $f$ defined over the interval $[a, b]$, the minimax approximation finds a polynomial $p(x)$ that minimizes $\underset{a\le x \le b}{max}|f(x)-p(x)|$.

%
%

\subsection{Chebyshev Approximation}
\label{subsec:chebyshev approx}

To approximate a continuous function $f$, defined over $[a, b]$, we first need to express $f$ as a series of Chebyshev polynomials at $[-1, 1]$. More precisely, $f$ is expressed as: $f(x) = \sum_{k=0}^{n}c_{k}T_{k}(x), \ \ x \in [-1, 1]$, where $c_{k}$ is the Chebyshev coefficient and $T_{k}(x)$ can be calculated from equation~\ref{equ:chebyshev}. As a next step, we calculate the coefficients of the polynomial and finally, express the polynomial in the original interval $[a, b]$. 
%
%

Our results for approximating the Sigmoid and ReLU activation functions, using Chebyshev approximation, are illustrated in table~\ref{tab:approximate sig and relu}. The approximation error for both activation functions is calculated using equation $E(x)= f(x) - p(x)$

\begin{table}[!ht]
	\small
	\caption{Approximating Sigmoid and ReLU}
	\label{tab:approximate sig and relu}
	\begin{tabular}{|l|l|l|l|l|}
		\hline
		\rowcolor{davy\'sgrey}		
		\multicolumn{5}{|l|}{\color{white}{\textbf{Approximation: Sigmoid}}}                \\ \hline
		\rowcolor{Gray}	
		x  & Interval    & Function(x) & Approximation & Difference \\ \hline
		-4 & {[}-5, 5{]} & 0.017986    & 0.016360      & -1.63e-03  \\ \hline
		-3 & {[}-5, 5{]} & 0.047426    & 0.049098      & 1.67e-03   \\ \hline
		-2 & {[}-5, 5{]} & 0.119203    & 0.118340      & -8.63e-04  \\ \hline
		-1 & {[}-5, 5{]} & 0.268941    & 0.268522      & -4.19e-04  \\ \hline
		1  & {[}-5, 5{]} & 0.731059    & 0.731478      & 4.19e-04   \\ \hline
		2  & {[}-5, 5{]} & 0.880797    & 0.881660      & 8.63e-04   \\ \hline
		3  & {[}-5, 5{]} & 0.952574    & 0.950902      & -1.67e-03  \\ \hline
		4  & {[}-5, 5{]} & 0.982014    & 0.983640      & 1.63e-03   \\ \hline
		\rowcolor{davy\'sgrey}				
		\multicolumn{5}{|l|}{\color{white}{\textbf{Approximation: ReLU}}}                   \\ \hline
		-4 & {[}-5, 5{]} & 0.000000    & -0.008871     & -8.87e-03  \\ \hline
		-3 & {[}-5, 5{]} & 0.000000    & 0.014340      & 1.43e-02   \\ \hline
		-2 & {[}-5, 5{]} & 0.000000    & -0.015085     & -1.51e-02  \\ \hline
		-1 & {[}-5, 5{]} & 0.000000    & -0.026883     & -2.69e-02  \\ \hline
		1  & {[}-5, 5{]} & 1.000000    & 0.973117      & -2.69e-02  \\ \hline
		2  & {[}-5, 5{]} & 2.000000    & 1.984915      & -1.51e-02  \\ \hline
		3  & {[}-5, 5{]} & 3.000000    & 3.014340      & 1.43e-02   \\ \hline
		4  & {[}-5, 5{]} & 4.000000    & 3.991129      & -8.87e-03  \\ \hline
	\end{tabular}
\end{table}

\section{Methodology}
\label{sec:methodology}

\begin{figure}[h]
	\centering
	\includegraphics[width=85mm,scale=2, frame=1pt]{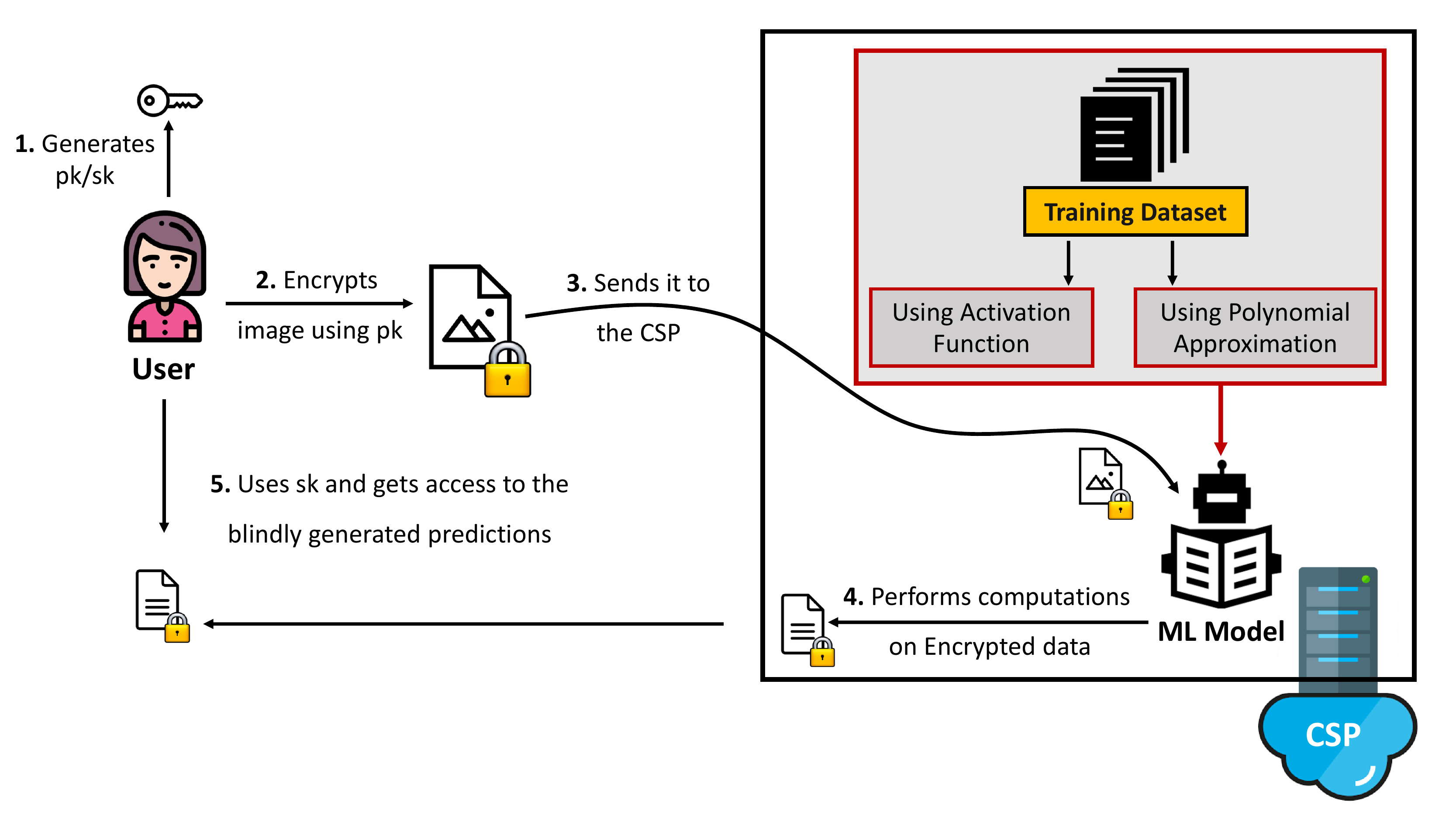}
	\caption{Blind Faith High Level Overview}
	\label{fig:client server model}
\end{figure}

We start this section by describing our system model. The topology of our work is illustrated in figure~\ref{fig:client server model}.  

In our model, we consider a CNN capable of analyzing large volumes of data (images) in a variety of domains. The CNN is deployed in a privacy-preserving manner in the CSP. 
To preserve the users privacy, we use HE. Using an HE scheme allows us to perform computations on encrypted data. However, HE schemes face certain limitations as they only support addition and multiplication operations. Most of the operations in a CNN are simple additions and multiplications and can thus be evaluated using HE. However, activation functions are non linear and as a result, we cannot use HE to perform operations on them. To this end, we replace the activation functions with polynomial approximations as already discussed in section~\ref{subsec:chebyshev approx}. While higher degree polynomials would provide us with a better approximation, they would also introduce higher computation and communication costs and hence, render our construction inefficient.

\paragraph*{\textbf{Flow}} The CNN model is deployed in a CSP and is trained using plaintext data. The weights and biases for this model are measured and made available to the CSP. For the training phase, we use the CNN given in figure~\ref{fig:convolutional training}. A user generates a public/private key pair for the HE scheme, encrypts an image and sends it to the CSP. Upon reception, the CSP runs the ML model and performs the classification in a privacy-preserving way. 
\begin{figure}
	\fbox{\begin{minipage}{24em}
			\begin{enumerate}
				\item \textit{\underline{Convolution}}: Input image 28 $\times$ 28, Window size 5 $\times$ 5, Stride(1,1), Number of input channels 1, number of output channels 5, Filters 5, Output 28 $\times$  28 $\times$  5.
				\item \textit{\underline{Activation Function}}: ReLU .
				\item \textit{\underline{Pooling}}: Mean, Window size 2 $\times$ 2 $\times$ 1, Stride(1,1), Output 14 $\times$ 14 $\times$ 5.
				\item \textit{\underline{Convolution}}: Window size 5 $\times$ 5, Stride(1,1), Filters 10, Output 14 $\times$ 14 $\times$ 10.
				\item \textit{\underline{Activation Function}}: ReLU .
				\item \textit{\underline{Pooling}}: Mean, Window size 2 $\times$ 2 $\times$ 1, Stride(1,1), Output 7 $\times$ 7 $\times$ 10.
				\item \textit{\underline{Fully Connected}}: Fully connects the incoming 7 $\times$ 7 $\times$ 10 nodes to the outgoing 128 nodes.
				\item \textit{\underline{Activation}}: ReLU 
				\item \textit{\underline{Fully Connected}}: Fully connects the incoming 128 nodes to the outgoing 10nodes.
				\item \textit{\underline{Softmax}}: Generate a probability for 10 nodes
			\end{enumerate}
	\end{minipage}}
	\caption{Convolutional Neural Network for Training Phase }
	\label{fig:convolutional training}
\end{figure}

\subsection{Inference Phase}
\label{subsec:Inference}
Although, the operations performed in the inference phase are nearly the same as in the training phase, there are few fundamental differences. For example, all operations in the inference phase are taking place on encrypted data while in contrast to the training phase where plaintext data is used. Similarly, the softmax which is part of the training phase is no longer available in the inference phase as shown in figure~\ref{fig:convolutional inference}. This operation can be skipped because it is a monotonically increasing feature, and the cipher texts from the output of the last completely connected layer can be compared to give the prediction.  Approximating or performing computation on encrypted data at this layer is out of the scope of the paper.

For the inference phase, we use the Fan-Vercauteren SHE scheme~\cite{fan2012somewhat}. 
The reason for using this specific scheme is that it allows us to perform \textit{both} addition and multiplication. Similarly, Single Instruction Multiple Data (SIMD) can be used to improve the performance of this scheme. Using SIMD technique, a batch of ciphertext is created in order to perform homomorphic evaluation in parallel. It is important to note that this scheme has three important parameters that affect the security level, and its performance: 

\begin{itemize}
	\item \underline{Polynomial Modulus}: This is an important parameter that affects the security level of the scheme. Polynomial modulus uses a power of two cyclotomic polynomial~\cite{thangadurai2000coefficients} and the recommended degrees for these polynomials are~1024, 2048, 4096, 8192 and beyond. On one side, a higher degree gives more security to the scheme while on the other side it degrades its performance.
	
	\item \underline{Coefficient Modulus}: This parameter determines the Noise Budget (NB) in the encrypted ciphertext. The coefficient modulus is directly proportional to NB and inversely proportional to the security level of the scheme.
	
	\item \underline{Plaintext Modulus}: The plaintext modulus affects NB in the freshly encrypted ciphertext. Additionally, it affects the NB consumption of homomorphic multiplications. For good performance, the recommendation is to keep the plaintext modulus as small as possible.
\end{itemize}

Each ciphertext in this encryption scheme has a specific quantity called NB -- measured in bits. The NB is determined by the above parameters and consumed by the homomorphic operations. The consumption of the NB is based on the chosen encryption parameters. For addition operations, this budget consumption is almost negligible in comparison to multiplication operation. In sequential multiplication that occurs at the convolutional and fully connected layer, the consumption of NB is very high. Hence, it is important to reduce the multiplicative depth of the circuit by considering appropriate encryption parameters. Once the NB drops to zero, then the decryption of ciphertext is not possible. Therefore it is necessary to choose the parameters to be large enough to avoid this, but not so large that it becomes ineffective and non functional.

While the HE scheme is based on polynomials, user's input is provided as a real number. Therefore, there is a clear mismatch between the two. Hence, it is important to use an encoding scheme that maps one to the other. To this end, the user encodes the input using the plaintext modulus and then encrypts it using the public key. The user also generates the encryption parameter and shares it with the CSP. To perform computations on the encrypted data, the CSP must have access to these parameters. 

Using the calculated weights and biases from the training phase and the encryption parameters, the CSP runs the inference phase on the encrypted image. The inference network is the same as the training network except that the activation functions are replaced by polynomial approximation and are built using an HE function. 

Now, the activation functions are substituted by polynomials. Since these polynomials only have addition and multiplication operations that are supported by HE. Consequently, we can perform encrypted computations on these functions. Similarly, the pooling operation in the inference phase is straightforward -- calculate the average value of four ciphertexts and multiply it with the appropriate values. However, the convolutional layer is a bit more expensive in terms of NB as it is a sequence of multiplication operations.

Additionally, the softmax layer is not a part of the inference network. Using the inference network, the CSP performs computation on encrypted data and obtains an encrypted output. The CSP does not have access to the secret key and thus cannot access the result. Furthermore, as the softmax layer is removed from the inference network the CSP is not able to predict the final output of the layer. 

At the end, the encrypted result of the output layer -- an array of 10 values which are homomorphically encrypted -- is sent back to the user. The user decrypts the results using the secret key and finds the output of the model which is the index corresponding to the highest among the 10 values.

\smallskip

At this point, it is important to highlight that the user utilized the ML model offered by the CSP and received the results without getting any valuable information about the underlying model. Similarly, the CSP ran the model on the encrypted image but at the same time was unable to extract any valuable information either for the content of the image or the actual prediction that sent back to the user. Hence, our model is considered as a privacy-preserving one. 

\begin{figure}
	\fbox{\begin{minipage}{24em}
			\begin{enumerate}
				\item \textit{\underline{Convolution}}: Input image 28 $\times$ 28, Window size 5 $\times$ 5, Stride(1,1), Number of input channels 1, number of output channels 5, Filters 5, Output 28 $\times$  28 $\times$  5.
				\item \textit{\underline{Activation Function}}: Approximated using polynomial approximation.
				\item \textit{\underline{Pooling}}: Mean, Window size 2 $\times$ 2 $\times$ 1, Stride(1,1), Output 14 $\times$ 14 $\times$ 5.
				\item \textit{\underline{Convolution}}: Window size 5 $\times$ 5, Stride(1,1), Filters 10, Output 14 $\times$ 14 $\times$ 10.
				\item \textit{\underline{Pooling}}: Mean, Window size 2 $\times$ 2 $\times$ 1, Stride(1,1), Output 7 $\times$ 7 $\times$ 10.
				\item \textit{\underline{Fully Connected}}: Fully connects the incoming 7 $\times$ 7 $\times$ 10 nodes to the outgoing 128 nodes.
				\item \textit{\underline{Activation}}: Approximated using polynomial approximation. 
				\item \textit{\underline{Fully Connected}}: Fully connects the incoming 128 nodes to the outgoing 10nodes.
			\end{enumerate}
	\end{minipage}}
	\caption{Convolutional Neural Network for Inference Phase }
	\label{fig:convolutional inference}
\end{figure}

\section{Performance Evaluation}
\label{sec:performance analysis}
In this section, we present our experimental results. 

\paragraph{\textbf{Experimental Setup}} 

All experiments were conducted in Python~3 using Ubuntu~18.04 LTS~64 bit (Intel Core~i7,~2.80 GHz,~32GB). For the training phase, we used Tensor flow\footnote{\url{https://www.tensorflow.org/}} to train our CNN model, while the actual experiments for that phase were conducted on Google Cola\footnote{\url{https://colab.research.google.com/}} (with GPU enabled).
Finally, for the inference phase we used Microsoft's Simple Encrypted Arithmetic Library (SEAL)~\cite{sealcrypto}.  

\paragraph{\textbf{Dataset}} To evaluate our model, similar to other works in the area, we used the MNIST dataset~\cite{lecun-mnisthandwrittendigit-2010} which consists of~60,000 images of handwritten digits. To train our CNN model we used 50,000 images while the rest 10,000 were used for testing. Each image is $28 \times 28$ pixel array and is represented by its gray level in the range of~0-255.

\subsection{Activation Function Approximation}
\label{subsec:ActiFuncApprox}

As we mentioned in the previous sections, in our approach we use Chebyshev polynomials to approximate the ReLU and Sigmoid activation functions where inputs are images encrypted with an SHE scheme. The polynomial approximation of the ReLU activation function is shown in table~\ref{tab:relu-polyapprox_degree}. Since the choices of the degree and the interval affect the performance of the model, it is necessary to choose suitable parameters. For this purpose, we conducted a series of experiments using different degrees and intervals\footnote{Due to space constraints, the results for approximating the ReLU and Sigmoid using degrees~7, 9 are only included.}. As can be seen in table~\ref{tab:relu-polyapprox_degree}, the activation functions are more accurately approximated when using polynomials of higher degree in small intervals. For example, the polynomial having degree~9 and interval $[-10, 10]$ more accurately approximate the ReLU function than the rest of the polynomials. 
The same applies to the Sigmoid activation function, where a high degree~9  and small interval $[-10, 10]$ give a better approximation as can be seen in~\ref{tab:sig-polyapprox_degree10}. 
However, the use of higher degree polynomials introduces a significant computation overhead, and small intervals limit the use of the approximation function. 

\begin{table}[!ht]
	
	\caption{Polynomial Approximation of the ReLU Function on Two Intervals ([-10,10], [-100,100]) using Different Degrees} 
	\label{tab:relu-polyapprox_degree}
	\scalebox{0.7}{
		\begin{tabular} {|p{7mm}|p{13mm}|p{45mm}|p{25mm}|}  \hline
			\rowcolor{Gray}
			Degree & Interval &Polynomial Approximation & ReLU Function \\ \hline
			
			%
			%
			7 &\(\displaystyle [-10, 10] \) & \(\displaystyle (-8.88178419700125e-21) \times x^7 + (3.66197231323541e-6) \times x^6 + \dots 
			\) 
			& \parbox[c]{1em}{
				\includegraphics[width=1in]{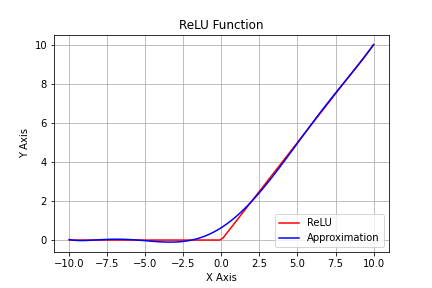}} \\ \hline 
			9 &\(\displaystyle [-10, 10] \) & \(\displaystyle (1.15960574476048e-21) \times x^9 - (7.03111115816643e-8) \times x^8 - \dots 
			\) & \parbox[c]{1em}{
				\includegraphics[width=1in]{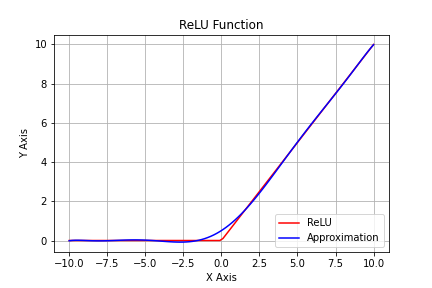}} \\ \hline 
						
			7 &\(\displaystyle [-100, 100] \) & \(\displaystyle (-6.82121026329696e-27) \times x^7 + (3.6619723132354e-11) \times x^6 + \dots
			\) & \parbox[c]{1em}{
				\includegraphics[width=1in]{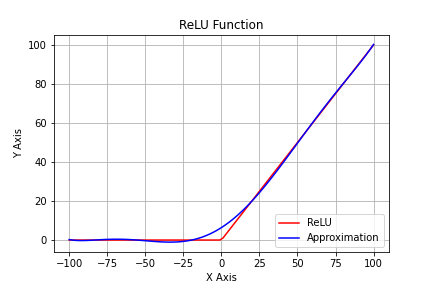}} \\
			\hline
			
			9 &\(\displaystyle [-100, 100] \) & \(\displaystyle (1.12777343019843e-29) \times x^9 - (7.03111115816644e-15) \times x^8 - \dots
			\) & \parbox[c]{1em}{
				\includegraphics[width=1in]{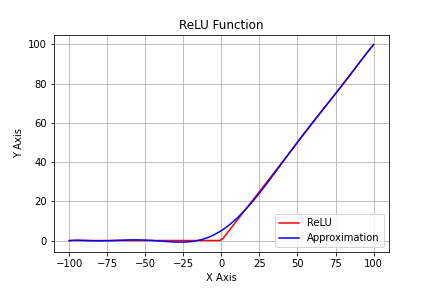}} \\
			\hline 
		\end{tabular}
	}
\end{table}

\begin{table}[h!]
	\caption{Polynomial Approximation of the Sigmoid Function on Two Intervals ([-10,10], [-100,100]) using Different Degrees} 
	\label{tab:sig-polyapprox_degree10}
	\scalebox{0.7}{
		\begin{tabular} {|p{7mm}|p{13mm}|p{45mm}|p{25mm}|}  \hline
			\rowcolor{Gray}	
			Degree & Interval &Polynomial Approximation & Sigmoid Function \\ \hline
			
			7 &\(\displaystyle [-10, 10] \) & \(\displaystyle (-4.34913635838155e-7) \times x^7 - (5.82079696725621e-16) \times x^6 + \dots\) & \parbox[c]{1em}{
				\includegraphics[width=1in]{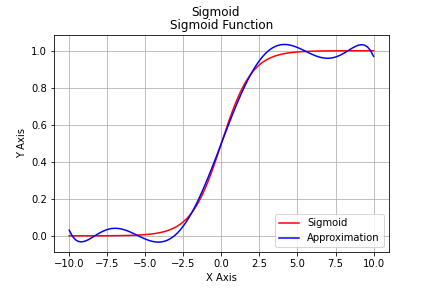}} \\ \hline 
			9 &\(\displaystyle [-10, 10] \) & \(\displaystyle (9.32721914680041e-9) \times x^9 + (1.39698499452418e-17) \times x^8 - \dots \) & \parbox[c]{1em}{
				\includegraphics[width=1in]{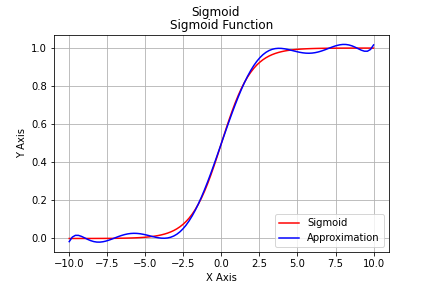}} \\ \hline 			
			
			7 &\(\displaystyle [-100, 100] \) & \(\displaystyle (-8.15672916212668e-14)\times x^7 - (5.82076636528339e-22)\times x^6 + \dots \) & \parbox[c]{1em}{
				\includegraphics[width=1in]{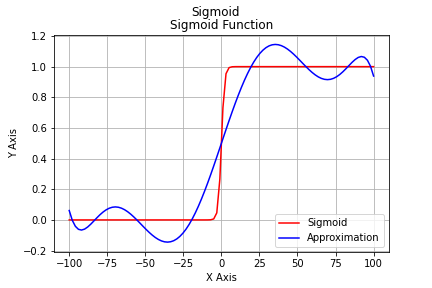}} \\
			\hline
			
			9 &\(\displaystyle [-100, 100] \) & \(\displaystyle (2.59190909648308e-17)\times x^9 + (1.3969822658824e-25)\times x^8 - \dots \) & \parbox[c]{1em}{
				\includegraphics[width=1in]{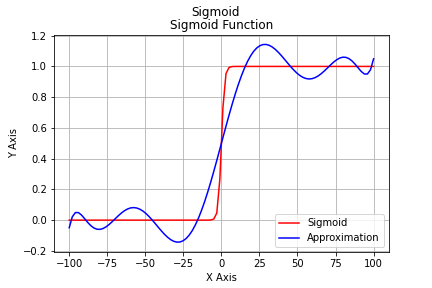}} \\
			\hline 
		\end{tabular}
	}
\end{table}

Furthermore, we performed a plethora of different experiments on the CNN model. We trained different networks by increasing the size of the convolutional layer and the size of the filters. We noticed that changing the number of convolutional layers and filters affects the overall accuracy of the network. As the size of the filter and layer increases, the accuracy of the network also increases. However, the efficiency of the network drops significantly. Hence, for the training phase, we considered the network given in figure~\ref{fig:convolutional training}. First, we trained the CNN model using the ReLU activation function. The measured accuracy for that part was~99.2\%. Then the same network was trained using the polynomial approximation function where we got an accuracy of~98.5\% -- a result that is very close to the original activation function.

For comparison, we used the model proposed in CryptoNets~\cite{gilad2016cryptonets} which is similar to the one proposed in our paper -- a convolutional layer, fully connected layers and an average polling layer as shown in table~\ref{tab:comp prev model}. Training the model with the ReLU activation function, the accuracy of our model was~99.20\% whereas CryptoNets achieved a~99\%. Similarly, for the approximated function we obtained an accuracy of~98.5\% while CryptoNets achieved~98.95\%. For the same network, the accuracy of the model proposed in~\cite{hesamifard2018privacy} was~99.02\% using the ReLU activation function and~99\% using the approximated function. 

\subsection{Performance Analysis of Performing Computation on Encrypted Data}
\label{subsec:HomoEvalCNN}
Now, we proceed by discussing how the use of homomorphic encryption can affect the performance of the NN model. In our work, we trained the CNN model on plaintext data while the classification was performed on the ciphertexts. As a result, we had to perform computations on two types of data -- plaintext and ciphertext. For this purpose, we used the SEAL library\footnote{\url{https://github.com/Lab41/PySEAL}} that allowed us to perform computations on ciphertext. Although the use of SEAL is straightforward, we still had to define certain parameters (see Section~\ref{subsec:Inference}).  

We performed a series of experiments using different encryption parameters. First, we looked at the polynomial modulus -- the encryption parameter used in SEAL. 
During the experiments we observed that a smaller value of polynomial modulus leads to a more efficient result but at the same time the accuracy is decreased. In contrast, a higher value of the polynomial modulus gives more accurate results, however, degrades the performance. The second encryption parameter is the coefficient modulus that decides the NB in the freshly encrypted ciphertext. This parameter is automatically set by SEAL based on the value of security level and polynomial modulus. Finally, increasing the value of the plaintext modulus, decreases the consumption of the NB.

\subsection{Comparison with the Existence Model}
\label{subsec:comparison with existence model}
Finally, we compared our results with state-of-the-art privacy-preserving NNs that utilize HE. The work proposed in CryptoNets~\cite{gilad2016cryptonets} is similar to ours. In CryptoNets, the model is trained on plaintext data and then the trained model is used for the classification of encrypted instances. In order to have a fair comparison, it is important to incorporate the same network used in both works. To this end, we used the CryptoNets model. 

Instead of using the overall performance of the model we decided to equate each layer. As can be seen in table~\ref{tab:comp prev model}, our model outperforms CryptoNets at both the encryption and decryption times as well as in the activation layer.

\begin{table}[!ht]
	\caption{Comparison with the Previous Models} 
	\label{tab:comp prev model}
	\begin{tabular}{|p{27mm}|p{25mm}|p{12mm}|p{8mm}|}
		\hline	
		\rowcolor{davy\'sgrey}		
		\color{white}\textbf{Layer} & \color{white}\textbf{Description} & \multicolumn{2}{c|}{\color{white}\textbf{Time}} \\  \hline
		\rowcolor{Gray}
		& & \color{darkgoldenrod}CryptoNets     & \color{darkgoldenrod}\TitleFull{} \\ \hline
		Encryption                                                           & Encoding+Encryption                                                                & 44.5        & 8.5242      \\ \hline
		1st convolution layer                                                & Same except the value of stride                                                    & 30          & 60.36       \\ \hline
		1st activation function                                              & \begin{tabular}[c]{@{}l@{}}Activation function\end{tabular} & 81          & 6.62        \\ \hline
		1st pooling layer                                                    & Mean pooling operation                                                             & 127         & 0.188       \\ \hline
		2nd convolutional layer                                              & -                                                                                  & -           & 64.822      \\ \hline
		2nd activation Function                                              & \begin{tabular}[c]{@{}l@{}}Activation function\end{tabular} & 10          & 0.199       \\ \hline
		2nd pooling layer                                                    & -                                                                                  & -           & 0.092       \\ \hline
		\begin{tabular}[c]{@{}l@{}}1st fully connected\\ layer\end{tabular}  & Generates 10 output                                                                & 1.6         & 12.1839     \\ \hline
		\begin{tabular}[c]{@{}l@{}}2nd fully connected \\ layer\end{tabular} & -                                                                                  & -           & 0.326       \\ \hline
		Decryption                                                           & Image decryption                                                                   & 3           & 0.0021      \\ \hline
	\end{tabular}
\end{table}

\paragraph{\textbf{Future Directions}} As a continuation of this work, we plan to explore the possibility of combining Functional Encryption (FE)~\cite{boneh2011functional}, with HE in an attempt to drastically improve the efficiency without sacrificing the accuracy of our construction. More precisely, we believe that it is possible to encrypt a user's data under an HE scheme, and then use a symmetric FE~\cite{bakas2021f,bakas2020multi} scheme in the inference phase of the CNN. Such an approach could significantly reduce the computational costs and the total running time of the algorithms.

\paragraph{\textbf{Open Science \& Reproducible Research}} 
To support open science and reproducible research as well as to give the opportunity to other researchers to use, test and hopefully extend/enhance our model we made all of our source code available both on Gitlab\footnote{\url{https://gitlab.com/nisec/blind_faith}} and Zenodo\footnote{\url{https://zenodo.org/deposit/5143787}}.

\section{Possible Societal Impact}
\label{sec:Impact}
Today's code created by machine learning is making surprisingly insightful moves, spotting previously undiscovered features in medical images, investing in shrewd trades on the stock market, composing music, writing articles, creating art, evaluating the effectiveness of teachers, deciding if an applicant is eligible for getting a loan and even choosing which students will be admitted into a college. Undoubtedly, ML models and their underlying applications are driving the big-data economy.

However, in practice, the systems using these models are incorporating proxies. It  has been observed that in many cases these proxies are bound to be inexact and often unfair. Considering that big data is not going away and predictive models are, and  increasingly will be, the tools we will rely on to run our institutions, deploy our resources and manage our lives, it is of paramount importance to bring transparency into the game. A possible solution would be to protect users' personal, in many cases unrelated to the application, data from a potentially unfair and biased ML model. Consequently, creating models capable of performing high accuracy predictions whilst being unable to learn anything about processed data or generated output could bring improved fairness to the age of data.  
With this work we hope to pave the way towards building digitally blind prediction/evaluation systems that will, by design, eliminate several proxies such as geography, gender, race, etc. and eventually have a tangible impact in building a fairer, democratic and unbiased societies. 

\bibliographystyle{ieeetr}
\balance
\bibliography{ISCC}

\end{document}